\documentclass[preprint,showpacs,prb]{revtex4}

\usepackage{graphicx}
\usepackage{dcolumn}
\usepackage{bm}

\begin{document}


\title{Orbital-Dependent Phase Control in Ca$_{2-x}$Sr$_x$RuO$_4$ ($0
\leq x \leq 0.5$)}

\author{Zhong Fang$^{1,2}$, Naoto Nagaosa$^{3,4}$, and Kiyoyuki
Terakura$^{5}$}

\affiliation{$^1$Spin Superstructure Project (SSS), ERATO, Japan
Science and Technology Corporation (JST), AIST Tsukuba Central 4,
Tsukuba 305-8562, Japan\\
$^2$Institute of Physics, Chinese Academy of Science, Beijing 100080, China\\
$^3$Correlated Electron Research Center
(CERC), AIST Tsukuba Central 4, Tsukuba 305-8562, Japan\\
$^4$Department of Applied Physics, University of Tokyo, Hongo 7-3-1,
Hongo, Tokyo 113-8656, Japan\\ 
$^5$Research Institute for
Computational Sciences (RICS), AIST Tsukuba Central 2, Tsukuba
305-8568, Japan}

\date{\today}

\begin{abstract}
We present first-principles studies on the orbital states of the
layered perovskites Ca$_{2-x}$Sr$_x$RuO$_4$. The crossover from
antiferromagnetic (AF) Mott insulator for $x < 0.2$ to nearly
ferromagnetic (FM) metal at $x=0.5$ is characterized by the systematic
change of the $xy$ orbital occupation. For the AF side ($x < 0.2$), we
present firm evidence for the $xy$ ferro-orbital ordering. It is found
that the degeneracy of $t_{2g}$ (or $e_g$) states is lifted robustly
due to the two-dimensional (2D) crystal-structure, even without the
Jahn-Teller distortion of RuO$_6$. This effect dominates, and the
cooperative occupation of $xy$ orbital is concluded. In contrast to
recent proposals, the resulting electronic structure explains well
both the observed X-ray absorption spectra and the double peak
structure of optical conductivity.  For the FM side ($x=0.5$),
however, the $xy$ orbital with half filling opens a pseudo-gap in the
FM state and contributes to the spin $S$=1/2 moment (rather than $S$=1
for $x$=0.0 case) dominantly, while $yz,zx$ states are itinerant with
very small spin polarization, explaining the recent neutron data
consistently.
\end{abstract}

\maketitle 

\section{Introduction}

The spin and orbital structures in Ca$_{2-x}$Sr$_x$RuO$_4$ (0$\leq x
\leq$0.5) attract much attention recently as a model system of
$t_{2g}$ electrons. It is an antiferromagnetic (AF) Mott insulator for
$x<$0.2, but nearly ferromagnetic (FM) metal at
$x$=0.5~\cite{Puchkov,Nakatsuji1,Braden1}. There are four electrons
occupying $t_{2g}$ states per Ru$^{4+}$, while $e_g$ states are
empty. The isovalent substitution of Sr for Ca does not change the
number of electrons, but modifies the crystal structure
systematically~\cite{Braden1} due to their different ionic radii.  The
key issue here is the orbital degree of freedom, which couples with
lattice and magnetism strongly. Extensive
experimental~\cite{Mizokawa,Lee,Kubota,Nakatsuji2,Braden2} and
theoretical~\cite{Fang1,Anisimov,Woods,Hotta} studies have been done
for this issue, while the present understanding of this compound is
quite controversial.

First, for the AF side ($x$=0.0), the preferential occupation of the
$xy$ orbital was predicted by earlier
calculations~\cite{Fang1,Anisimov}. However the O 1s X-ray absorption
spectroscopy (XAS) study reported the occupation of about 0.5 holes in
the $xy$ orbital\cite{Mizokawa}. In addition to this, the
photoemission experiment suggested strong spin-orbit coupling (SOC),
leading to the proposal of complex $t_{2g}$
orbitals~\cite{Mizokawa}. The presence of 0.5 holes in the $xy$
orbital also led Hotta {\it et al.}'s~\cite{Hotta} to propose the
``antiferro-orbital ordering'' (AFO) with 2x2 periodicity of RuO$_6$
in the plane. The recent experiment on optical conductivity was
interpreted in terms of this orbital pattern~\cite{Lee}.  However, the
lattice distortion corresponding to the AFO (i.e., the 2x2 structure)
is not observed experimentally, and the orbital pattern of AFO is not
consistent with the recent resonant X-ray scattering (RXS)
experiment~\cite{Kubota}, in which no superlattice peak was observed.
Second, for the nearly FM side ($x$=0.5), the fitting of the
Curie-Weiss type susceptibility suggested $S$=1/2 spin
moment~\cite{Nakatsuji1}, in sharp contrast with the high-spin ($S$=1)
state for $x$=0.0.  A similar picture may hold also for
$x$=0.2~\cite{Nakatsuji2}. The recent polarized neutron diffraction
studies for $x$=0.5~\cite{Braden2} suggested dominant magnetization
distribution on the $xy$ orbital rather than the theoretically
suggested $yz/zx$ orbitals~\cite{Anisimov}.  Therefore it is an
important and challenging issue to determine the orbital state of
Ca$_{2-x}$Sr$_x$RuO$_4$ explaining consistently the available
experiments, which we undertake in this paper.

We show that the crossover of magnetic properties with increasing
doping $x$ is characterized by the systematic change in the $xy$
orbital occupation. For the AF side ($x<$0.2), we conclude that the
predicted ferro-orbital (FO) ordering with dominant $xy$ occupation
explains well (a) the XAS spectra, and (b) the double peak structure
of the optical conductivity. We further point out that the
stabilization of the $xy$ state is quite robust, and mostly due to the
two-dimensional (2D) crystal-field, which has not been considered
seriously so far.  In contrast to the AF side, for $x$=0.5, the
electronic structure is characterized with the $yz/zx$ states with
small moment, and the $xy$ orbital with half filling, which opens a
pseudo-gap in the FM state and contributes to the $S$=1/2 moment
dominantly. This picture, which is in strong contradiction to the
previous proposal~\cite{Anisimov}, explains the neutron
data~\cite{Braden2} consistently.

\section{Calculation Method}

The calculations were done with the first-principles plane-wave
pseudopotential method.  In our previous work~\cite{Fang1}, the
effects of structural distortions were emphasized by taking into
account important distortions mostly, however here we use the
experimental structures to compare the calculated results with the
measured spectra. Five experimental structures are considered, i.e.,
$x$=0.0 at 11K and 295K, $x$=0.1, 0.2 and 0.5 at
10K~\cite{Braden1}. Two kinds of magnetic states, FM and staggered AF
states, are considered for each structure.  In the previous
work~\cite{Fang1}, the local spin density approximation (LSDA) was
adopted for the entire range.  The validity of LSDA for predicting the
stable magnetic state of ruthenates was
demonstrated~\cite{MazinSingh}.  However for Ca$_2$RuO$_4$, though
LSDA can predict the stability of AF magnetic order, the band gap does
not open~\cite{Fang1,Woods}.  As we aim to analyze the optical
conductivity also, a proper reproduction of the band gap is important.
Therefore, we employ the LDA+$U$ scheme~\cite{Solovyev,JPC} for the
insulating region (0$\leq x < $0.2) with $U_{\rm eff}$=2.5eV to
reproduce the band gap of Ca$_2$RuO$_4$.  The LSDA was employed for
other region of $x$.  The details of other aspects of the calculation
were described in the previous paper~\cite{Fang1}. For the studies of
XAS spectra and optical conductivity, the interband transitions are
calculated from the converged Kohn-Sham wave functions and eigen
values, by using the core compensation form of pseudo-wave functions
(see Ref.~\cite{Fang2} for technical details).

\section{Results and Discussion}

\subsection{Ca$_2$RuO$_4$ ($x=0$)}
We start with Ca$_2$RuO$_4$ at 10K. Our calculations predict the AF
ground state with energy gain about 28.2meV with respect to the FM
state.  The mechanism for stabilizing the AF state was analyzed in our
previous paper~\cite{Fang1} in terms of lattice distortion,
flattening, rotation and tilting of RuO$_6$.  The projected density of
states (PDOS) in Fig.1(a) suggests that the present situation may
correspond to the localized spin picture of $S$=1, where the $xy$
states are fully occupied and the $yz/zx$ orbitals are half filled.
In reality, as the Ru $4d$ orbitals are extended, inter-site
hybridization modifies the picture quantitatively.  For example, the
occupation numbers (in LDA+U) are given as follows:
$n^\uparrow_{xy}$=0.86, $n^\uparrow_{yz/zx}$=0.87,
$n^{\downarrow}_{xy}$=0.79 and $n^{\downarrow}_{yz/zx}$=0.28, with up
and down arrows indicating the spins (see Figs.1(a) and 2(a)). The
cooperative occupation of the $xy$ orbital for all Ru sites forms the
FO ordering.  These occupation numbers give 1.25$\mu_B$ as the
magnetic moment of a Ru atom $M_{Ru\_total}$, which agrees well with
the experimental value of 1.3$\mu_B$~\cite{Braden1}.  As for the spin
polarization of oxygen, the in-plane oxygen atoms have no net
polarization due to symmetry while the apical oxygen atoms are
strongly polarized (about 0.1 $mu_B$) due to the strong $p-d$
hybridization because the $yz/zx$ orbitals which extend toward apical
oxygen contribute dominantly to the Ru spin polarization.  The
corresponding values in LSDA for the above orbital population are
$n^\uparrow_{xy}$=0.80, $n^\uparrow_{yz/zx}$=0.82,
$n^{\downarrow}_{xy}$=0.71 and $n^{\downarrow}_{yz/zx}$=0.40.
Therefore, the basic feature of the orbital population remains the
same in LSDA, though the orbital polarization and the spin
polarization are reduced compared with those in LDA+$U$.  For example,
LSDA gives 0.93$\mu_B$ as $M_{Ru\_total}$.  The reduction in these
polarizations is due to the stronger inter-site hybridization caused
by the vanishing band gap in LSDA.  Therefore, we believe that the
results obtained by LDA+U which adjusts the effective Coulomb
repulsion $U_{\rm eff}$ to reproduce the band gap should be
quantitatively more reliable than those by LSDA.  The present results
for the electronic structure are consistent with previous
calculations~\cite{Anisimov,Woods}.

As for the orbital character, there are three factors mainly
contributing to the stabilization of the $xy$ orbital:

1) The energy level splitting due to 2D crystal field in the layered
structures (in contrast to its three-dimensional (3D) counterpart,
where the $t_{2g}$ (or $e_g$) states are degenerate). Fig.3 (main
panel) shows the calculated PDOS for the non-magnetic (NM) state of
Ca$_2$RuO$_4$ without the Jahn-Teller distortion (i.e. three Ru-O
bonds having equal length). In the inset of Fig.3, we show the
calculated energy difference between the $xy$ and the $yz,zx$ band as
a function of the tetragonal JT distortion, by calculating the center
of gravity for each band. Clearly, even for the hypothetical
tetragonal Ca$_2$RuO$_4$ without JT distortion, the $xy$ orbital is
lower in energy than the $yz$ and $zx$ orbitals by about 0.2eV as
estimated by our calculation. This lowering of $xy$ state energy is
added to that due to the tetragonal distortion as demonstrated in the
almost uniform downward shift of the dashed line (Ca$_2$RuO$_4$) from
the solid line (CaRuO$_3$) in the inset of Fig.3.  As a first
approximation, the origin of the energy level splitting is attributed
to the following geometrical aspect in the second neighbor
configurations: in Ca$_2$RuO$_4$, Ru-O-Ru in the $ab$ plane is
replaced with Ru-O-Ca along the $c$ axis.  Two factors coming from
this geometrical aspect contribute to the energy level splitting.
First, the electrostatic potential due to the reduced positive charge
of the second neighbor Ca$^{2+}$ along the $c$ axis will raise the
energies of those orbitals extending along the $c$ axis, like the $yz$
and $zx$ (or $3z^2-r^2$) states. This can be seen in Fig.3 from the
global upward shift of $yz, zx$ bands and their oxygen bonding
counterparts relative to those of the $xy$ state. Second, due to the
absence of Ru-O bond at one side of the apical oxygen along the $c$
axis, the remaining Ru-O bonds are strengthened. This will further
push up $yz$ and $zx$ states, which are the anti-bonding parts of the
$2p$-$4d$ hybridization.

2) The compressive JT distortion. The $xy$ state is further lowered by
about 0.06eV (see inset of Fig.3) due to the 2\% shrinkage of the
apical Ru-O bond length observed for Ca$_2$RuO$_4$ at 10K.

3) The orbital-dependent hybridization. In the AF state, the
inter-site hybridization between the occupied orbitals and the
unoccupied ones, which is the origin of the super-exchange, will push
up the unoccupied $yz/zx$ states, and again enhance the splitting
between $xy$ and $yz/zx$ in minority spin by about 0.07eV from our
calculations. This effect does not exist in the FM state, and will
explain the reduced occupation of $xy$ orbital in the FM solution even
for $x$=0.0 (see Fig. 1(b)).

As the results, $n^{\downarrow}_{xy}$ reads 0.79, 0.72, and 0.67 for
the cases corresponding to ($x$=0.0, 11K, AF), ($x$=0.1, 10K, AF: JT
distortion is nearly vanishing), and ($x$=0.0, 11K, FM), respectively.
It is clear that the $xy$ occupation is dominantly determined by the
energy level splitting due to the 2D structure, which is comparable
with the typical band width of $t_{2g}$ states. The existence of such
2D crystal-field is common for all the layered perovskites, while it
has not been considered seriously so far.  This effect, which is not
taken into account in the analysis by Hotta and Dagotto~\cite{Hotta},
will certainly suppress the stability of their proposed AFO state for
Ca$_2$RuO$_4$.

We also performed self-consistent calculations by including the SOC
using the relativistic fully separable pseudopotentials~\cite{SOC} in
the framework of non-collinear magnetism. We found that the obtained
orbital occupations are almost identical to those for the case without
SOC, in the AF gound state of Ca$_2$RuO$_4$. This is consistent with
above discussions in the sense that the crystal field is strong enough
to quench the orbital moment and stabilize the real-orbital $xy$.
Note that the photoemission measurement~\cite{Mizokawa} was done at
150K which is above the N\'{e}el temperature where the degree of
orbital polarization is reduced.

The electronic structures obtained in the present work can account
well also for the observed XAS~\cite{Mizokawa} for Ca$_2$RuO$_4$
though the number of holes in the $xy$ state is not exactly 0.5 (the
estimated number of holes is between 0.2 and 0.3 from our calculations
for the AF ground state).  There are two important aspects in the
experimental analysis.  One is the distribution of holes among
$t_{2g}$ states and the other is its variation from the
low-temperature AF phase to the high-temperature paramagnetic (PM)
phase.  Such information was derived from the observed relative
intensity of two XAS peaks (A and B) as a function of light incidence
angle $\theta$.  The peaks A and B correspond to the $1s$-$2p$
transition at the apical and the in-plane oxygen, respectively.  In
the single electron transition limit, we obtain the XAS spectra from
the converged Kohn-Sham eigen states by calculating the matrix
elements of the optical transition~\cite{Fang3}.  For the
low-temperature AF phase, the calculation is straightforward.
However, we have to introduce an approximate treatment for simulating
the XAS in the high-temperature PM state.  Here we use the following
simplest approximation in the cluster expansion technique for treating
the alloy problem~\cite{Connolly}, assuming that a magnetic state can
be regarded as an alloy with two constituents, up spin state and down
spin state.  As the simplest approximation, we use a dimer for the
cluster.  Then the XAS of paramagnetic state can be approximated
simply as an average of XAS of FM and AF states.  (If we use a larger
cluster, other magnetic orders have to be taken into account and the
short-range order effect can be included.)  As shown in Figs.4(a) and
(c), our calculated XAS at 11K and 295K can be well compared with the
experimental spectra taken at 90K and 300K (Fig.3 of
Ref.~\cite{Mizokawa}). At elevated temperature, which tends to
suppress the compressive JT distortion and enhance the spin
disordering (increasing FM component in spin configurations), the
$n^{\downarrow}_{xy}$ will decrease as obtained in our calculations
for the 295K structure (see Figs.1(c) and (d)), being consistent with
the experimental tendency.

Recently, the anisotropic optical conductivities in Ca$_2$RuO$_4$ were
measured by Lee {\it et al.}~\cite{Lee} and Jung {\it et
al.}~\cite{Jung}. For the {\bf E}//$a$ spectra (see Fig.5), two peak
structures (called $\alpha$ and $\beta$ at about 1.0 eV and 2.0 eV
respectively) are related to the $4d-4d$ transitions (the strong peak
at 3.0eV is due to the $p-d$ charge transfer). By increasing the
temperature, spectral weight transfer from the strong peak $\beta$ to
the weak peak $\alpha$ is observed and simultaneously the gap is
reduced. We calculate~\cite{Fang2} the optical conductivities by
using the Kubo formula~\cite{CallawayWang}, and all transition matrix
elements are calculated from first-principles~\cite{Fang2}. As shown
in Fig.5, the experimental results can be well explained by our
calculations, and the corresponding transition paths are indicated in
Fig.1 by red arrows. Again we use the simple average of FM and AF
solutions to simulate the PM state at 295K. For the low temperature AF
state, the peak $\beta$ at 2.0 eV is mostly due to the transition from
the occupied majority spin $yz$ ($zx$) state at one Ru$^{4+}$ site to
the unoccupied minority spin $yz$ ($zx$) state at the neighboring
Ru$^{4+}$ site.  In the simple energy level picture (i.e, the
occupation number of each orbital is regarded as integers, either 0 or
1), it is easy to understand the origin of peak $\beta$ (which
corresponds to transition shown in Fig.1(c) of Ref.~\cite{Lee} with
energy $U+J$), but not that of $\alpha$. This argument has been used
by Lee et al.~\cite{Lee} against the FO ordering. So they pursue the
proposal for the AFO state~\cite{Hotta}, which has been shown in the
above discussions not to be favored. However, our calculation clearly
shows the existence of peak $\alpha$ at about 1.0eV, which is due to
the non-vanishing transition matrix element between $xy$ and $yz/zx$
states, and crucially depends on the orthorhombic distortion and the
admixture of $xy$ and $yz/zx$ states of Ru-$4d$. One of the strong
contributions to $\alpha$ is the transition from the occupied majority
spin $xy$ state at one Ru$^{4+}$ site to the unoccupied minority spin
$xy$ component mixed into the $yz, zx$ states at the neighboring
Ru$^{4+}$ site (see Fig.1(a)).  Therefore, the double peak structure
is just the natural consequence of strong hybridization.  The peak
$\beta$ should be strong at the low-temperature AF phase, and
sensitive to the magnetic ordering. Actually, this transition should
be suppressed in the FM configuration (as shown in Fig.1(d)), because
of the flipping of the spin.  The peak $\alpha$ is weak at low
temperature AF state, and should be sensitive to the occupation number
of $xy$ orbital. The less $n_{xy}$ is, the stronger the transition
is. Actually, this transition is much enhanced in the FM configuration
as shown in Fig.1(d). Therefore, with the increasing temperature, the
enhancement of peak $\alpha$ can be understood by the increasing
apical Ru-O bond length and the increasing FM components in the spin
configuration, both of which will suppress the $xy$ occupation, while
the reduction of peak $\beta$ is mostly due to the spin
disordering. The reduction of the gap at elevated temperature can be
understood in terms of reduction in the AF long-range order.

\subsection{Ca$_{2-x}$Sr$_x$RuO$_4$}

Above results present firm evidence for the $xy$ FO ordering for
$x$=0.0. Now going from $x$=0.0 to 0.1, we see quantitative change of
the electronic states (Fig.1(e) and Fig.2), while the ordering pattern
does not change qualitatively.  The elongation of the Ru-O bond along
the $c$-axis~\cite{Braden1} (going from $x$=0.0 to 0.1) tends to
suppress the occupation of $xy$ orbital, and to destroy the AF
order. As a result, we see from Fig.2 the followings: 1) the energy
gain of AF state with respect to the FM state reduces, being
consistent with the reduced N\'{e}el temperature $T_{N}$ observed
experimentally; 2) the reduction of ordered moments; 3) electron
transfer in the minority spin channel from the $xy$ orbital
(n$^{\downarrow}_{xy}$) to the $yz/zx$ orbitals
(n$^{\downarrow}_{yz/zx}$); 4) redistribution of magnetization from
the $yz/zx$ orbitals ($M_{yz/zx}$) to the $xy$ orbital ($M_{xy}$).
The reduction of n$^{\downarrow}_{xy}$ is also visible from the
calculated XAS spectra (shown in Fig.4(b)), where the relative
intensity of two peaks (A and B) is just between those of Figs.4(a)
and (c).

However for $x$=0.5, where RuO$_6$ is quite elongated, and has only
strong rotation around the $c$-axis without tilting, the situation is
completely different (Fig.1(f) and Fig.2).  These structural
modifications produce basically two important changes in the
electronic structure compared with the case of Ca$_2$RuO$_4$: 1)
reduction in $n^{\downarrow}_{xy}$ due to the elongation; 2)
broadening of the $yz/zx$ bands due to the absence of
tilting. Nevertheless, the existence of RuO$_6$ rotation keeps the
$xy$ band narrow~\cite{com1}. This will contribute to the high density
of states at the Fermi level of non-magnetic solution, leading to the
Stoner-type FM instability. Since this instability mostly comes from
the $xy$ state, we observe from the FM solution (Fig.1(f)) that the
$xy$ states are strongly spin polarized, opening a pseudo-gap, and
contributing to the magnetization dominantly (Fig.2).  On the other
hand, $yz/zx$ bands are quite broad, and are located around the Fermi
level with much reduced spin polarization compared with the case of
$x$=0.0.  In contrast to the $S=1$ picture with the magnetic moment
supported by the $yz/zx$ orbitals for the $x$=0 case (Ca$_2$RuO$_4$),
the present case may correspond to the $S$=1/2 picture where the $xy$
orbital contributes to the magnetic moment.  This picture accounts
well for the observations by the Curie-Weiss fitting with $S$=1/2 of
susceptibility~\cite{Nakatsuji1} and the polarized
neutron~\cite{Braden2} showing the spatial magnetic moment
distribution of the $xy$ character. The large spin polarization for
the in-plane oxygen atoms and the negligibly small one for the apical
oxygen atoms (Fig.2(c)) are the results coming from the $xy$ orbital
origin of the Ru magnetic moment and are consistent with the
experiment~\cite{Braden2}.

In contrast to the present study, the proposal in
Ref.~\cite{Anisimov}, which claims the contribution of $yz/zx$ states
to the magnetization, was obtained by neglecting the RuO$_6$ rotation,
which is an important ingredient for obtaining our orbital-dependent
picture. In their treatment~\cite{Anisimov}, an elaborate LDA+DMFT
(dynamical mean field theory) scheme was used, but high symmetry
structure of Sr$_2$RuO$_4$ was assumed.  Then the effects of doping
are simulated by increasing the size of $U$, instead of changing the
band-width. However, we pointed out in our calculation that the strong
RuO$_6$ rotation will reduce the band width of $xy$ significantly, but
not that of $yz,zx$ bands. This difference produces the main source of
the discrepancy discussed above. The characteristic results of our
picture for $x$=0.5 are the much suppressed (enhanced) peak A (B) in
the XPS spectra (Fig.4(d)), and the structureless optical spectra
below the charge-transfer peak (Fig.5(b)). These predictions should be
confirmed by further experiments.

Our picture for $x$=0.5 was obtained by the LSDA calculations which
tends to overestimate the FM stability. Experimentally no FM long
range order has been observed, while the significant enhancement of
susceptibility down to 2K~\cite{Nakatsuji1} clearly suggested the
existence of strong FM correlation (at least for short
range). Actually, the recent experiments down to
0.3K~\cite{Nakatsuji2} suggested the existence of weak FM component
(cluster glass) at this doping.  This may suggest that some magnetic
orders may be energetically in near degereracy with FM state.  At
present we do not have any clear idea about what magnetic orders they
may be and the problem is left for future studies.

Finally, we discuss only briefly the case for the critical point
$x$=0.2.  10K is already above the metal-insulator transition
temperature~\cite{Braden1} in this case, and the system crystallizes
in the $L-Pbca$ phase with a long apical Ru-O bond, suggesting the
adjacency to the metallic side ($x$=0.5).  However, in contrast to the
case of $x$=0.5, the tilting of RuO$_6$ still exists to reduce the
width of $yz/zx$ bands as well at $x$=0.2.  This aspect tends to
stabilize the insulating AF state.  If we adopt LSDA, the former
aspect is emphasized to make the case of $x$=0.2 very similar to the
case of $x$=0.5 as can be seen in Fig.2.  On the other hand, the
LDA+$U$ method with $U_{\rm eff}=$2.5 eV emphasizes the latter aspect
to bring the AF state very close to the FM state with the energy
difference given by $E_{\rm FM} - E_{\rm AF}=-3$ meV.  Note, however,
that the similar LDA+$U$ calculation for $x$=0.5 enhances the
stability of the FM state with respect to the AF state and doubles the
energy difference in Fig.2(d).  These calculations, though not
conclusive, clearly suggest that the system with $x$=0.2 may be in the
critical situation.

\section{Concluding Remarks}

We have presented a systematic picture for the orbital states in
Ca$_{2-x}$Sr$_x$RuO$_4$ ($0 \leq x \leq 0.5$).  For the AF side ($x <
$0.2), we conclude that the orbital ordering is of the FO type with
the dominant $xy$ occupation and that the $yz/zx$ orbitals contribute
to the magnetic moment.  The electronic structure corresponds
basically to $S$=1 in the localized spin picture.  However, strong
intersite hybridization due to extended Ru $4d$ orbitals significantly
modifies the magnetic moment distribution and the present calculation
agrees well with the experiments.  For $x$=0.5, our LSDA calculation
predicts the system to be ferromagnetic and the narrow $xy$ band with
half filling may correspond to an $S$=1/2 localized spin picture.
Experimentally the system shows strong tendency toward ferromagnetism
but remains paramagnetic for $x$=0.5 even down to very low
temperature.  Only below 0.3K, the existence of cluster glass phase
with weak FM component was observed.  The LSDA calculation seems to
overestimate the stability of ferromagnetism in other cases also.
This problem is left for future studies.  However, the present result
for the character of the spin-density distribution is consistent with
the neutron data.  We emphasized that the structural modifications
such as rotation, tilting and flattening of RuO$_6$ octahedron have to
be fully taken into account in order to describe properly the
systematic variation of the orbital states.  Our results can explain
other existing experiments, for example, XAS and optical conductivity,
quite consistently.  Furthermore, we made predictions for these
experiments for the $x$=0.5 compound, which await experimental tests.

\begin{acknowledgments}
The authors thank Prof. Y. Tokura, Dr. J. H. Jung and Dr. M. Kubota
for fruitful discussions and providing their experimental data. One of
the authors (Z.F.) acknowledges supports from NSF of China
(No.90303022).
\end{acknowledgments}

\newpage
\begin{figure}[t]
\includegraphics[scale=0.85]{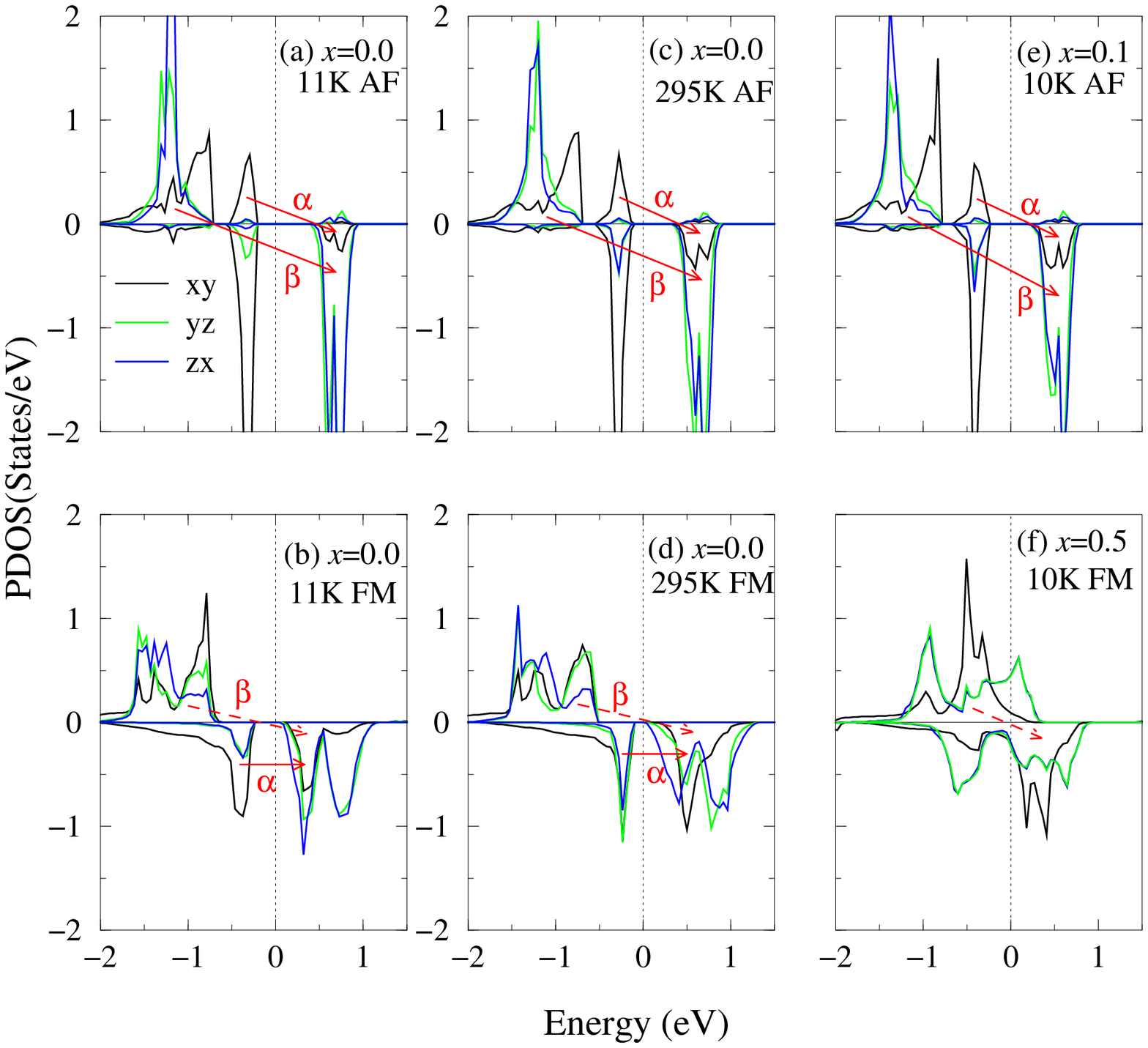}
\caption{The calculated PDOS of Ru-4$d$-$t_{2g}$ orbitals for various
states. The solid red arrows indicate the main optical transition
paths, while the dashed red arrow shows the transition which is
significantly suppressed. Note the transitions indicated here mean
that from the states at one Ru$^{4+}$ site to the corresponding states
at nearest-neighboring Ru$^{4+}$ site.}
\end{figure}

\newpage
\begin{figure}
\includegraphics[scale=0.9]{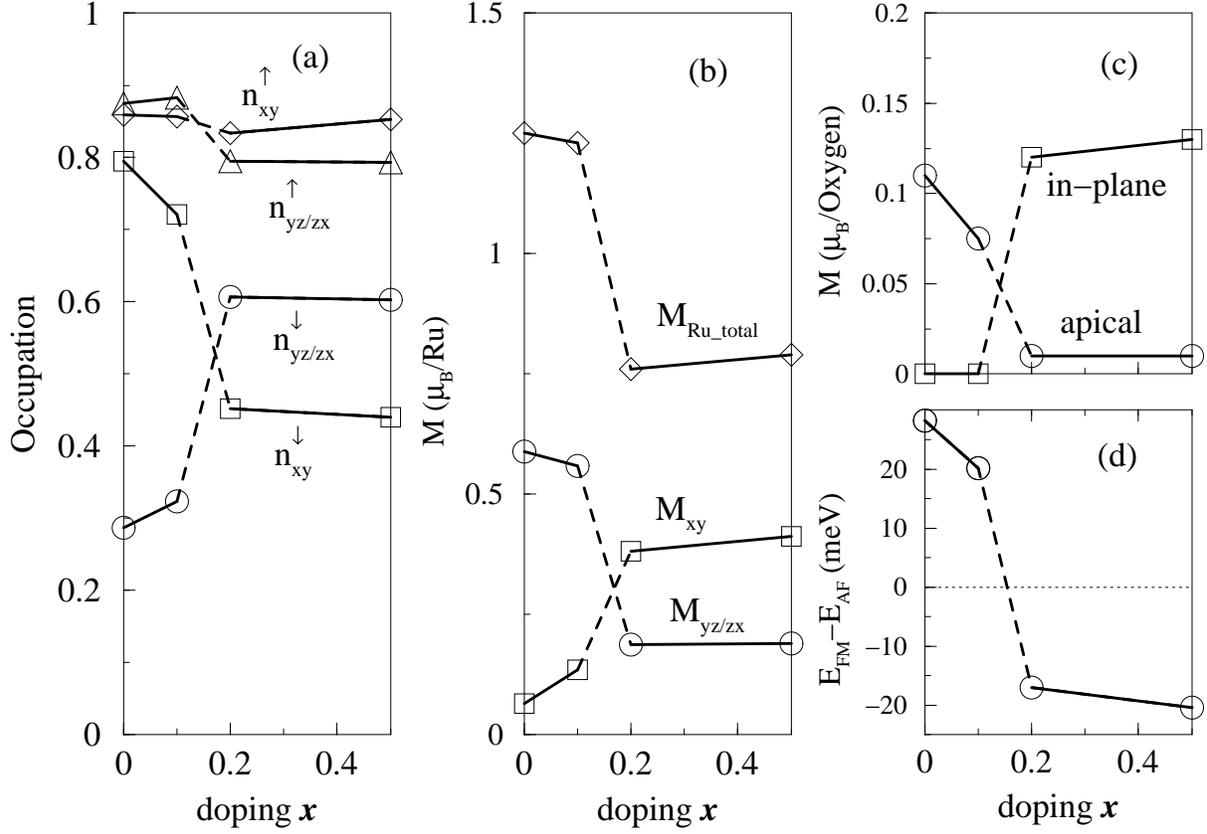}
\caption{The calculated (a) orbital occupation numbers and (b)
magnetic moments for the ground state of Ca$_{2-x}$Sr$_x$RuO$_4$ with
different doping $x$.  $M_{xy}$ ($M_{yz/zx}$) denotes the magnetic
moment associated with the $xy$ ($yz/zx$) orbital and
$M_{Ru\_total}=M_{xy}+2.0 M_{yz/zx}$.  The absolute value of magnetic
moments from oxygens are shown in (c) (see text for further
explanations). The panel (d) shows the total energy difference between
the FM and AF states. The AF side ($x$=0.0, 0.1) is connected with the
FM side ($x$=0.2, 0.5) by dashed lines.}
\end{figure}

\newpage
\begin{figure}
\includegraphics[scale=0.9]{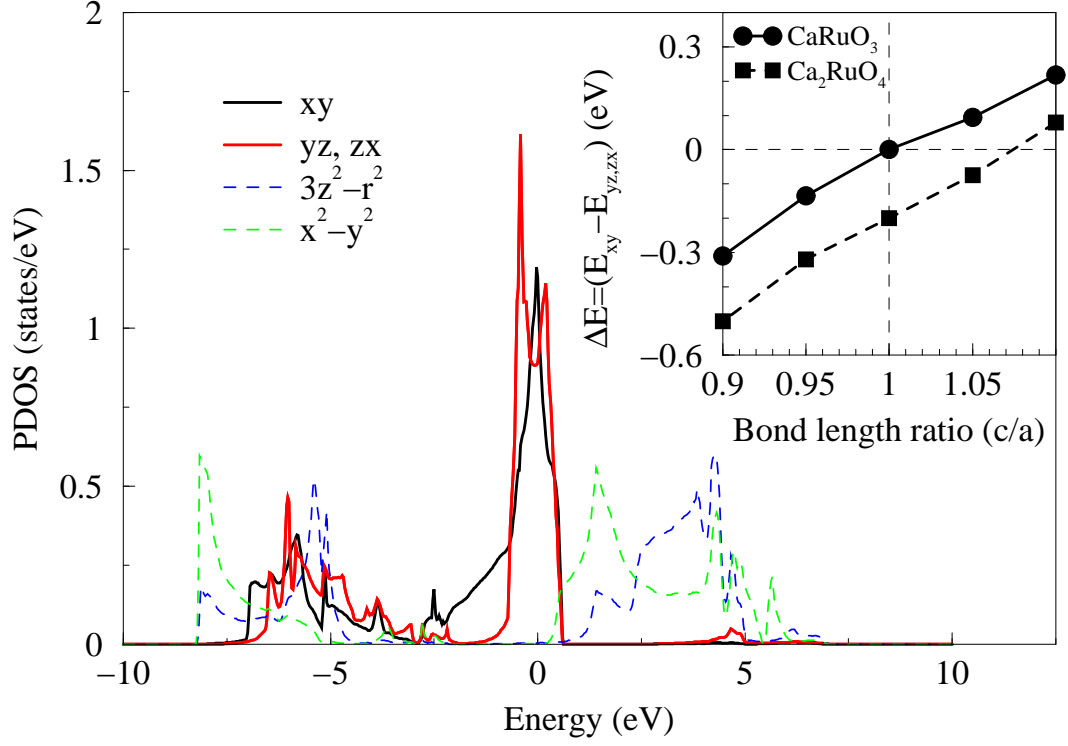}
\caption{The projected density of states of tetragonal Ca$_2$RuO$_4$
without the JT distortion in the NM state. The inset shows the
calculated energy level splitting between the $xy$ and the $yz, zx$
states for CaRuO$_3$ and Ca$_2$RuO$_4$ with tetragonal Jahn-Teller
distortion, which is defined as the bond length ratios (Ru-O bond
length along the $c$-axis versus that in the $ab$-plane). }
\end{figure}

\newpage
\begin{figure}
\includegraphics[scale=0.9]{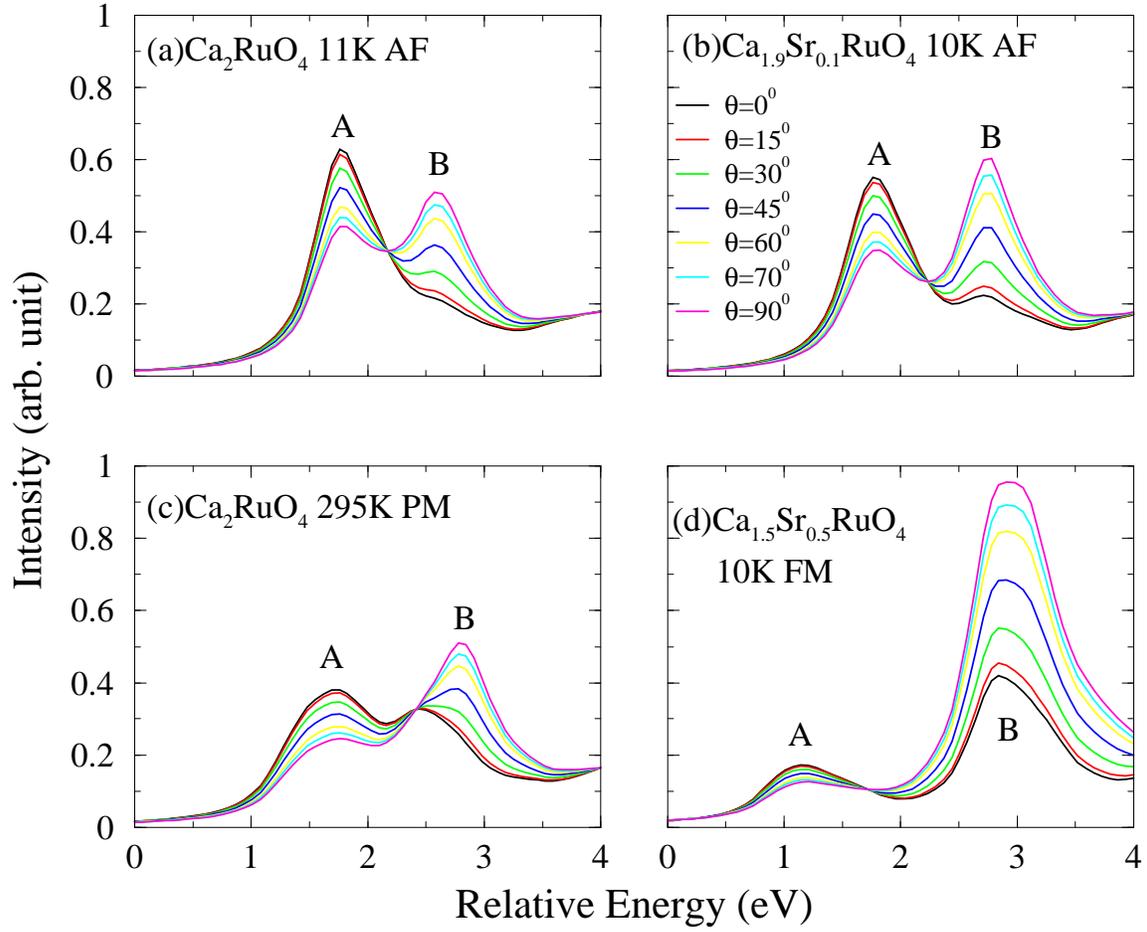}
\caption{The calculated relative intensity of two XAS peaks (A and B
located at 528.5 and 529.5 eV experimentally) as a function of light
incidence angle $\theta$. }
\end{figure}

\newpage
\begin{figure}
\includegraphics[scale=1.0]{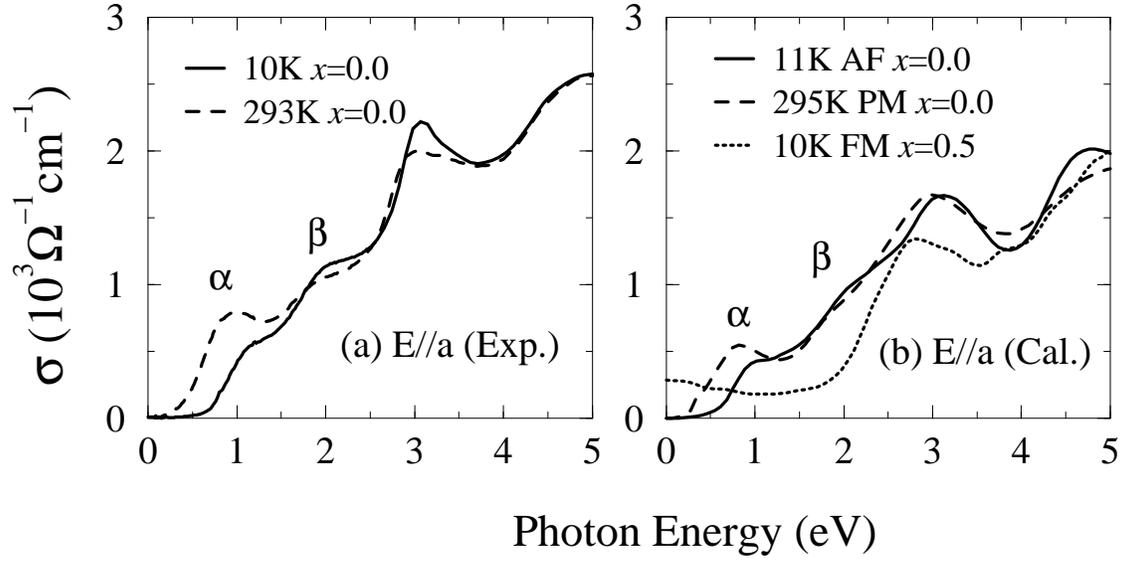}
\caption{The (a) experimental~\cite{Jung} and (b) calculated optical
conductivities. The interband transitions are calculated by using the
Kubo formula. The simple average of FM and AF solutions is used to
simulate the PM state at 295K.}
\end{figure}


\begin{thebibliography}{}
\bibitem{Fang3} The calculation was done for the oxygen $2s$ level
instead of the $1s$ core as the initial state. Since the $2s$ state
is already well below the Fermi level, the difference with $1s$ should
be small.
\bibitem{com1} The $xy$ band is still broad in the sense that its PDOS
has a long tail in the lower energy range.  However, the peak
structure of PDOS around the Fermi level for the non-magnetic state is
very sharp with the RuO$_6$ rotation as clearly shown in Fig.2(b) of
Ref.~\cite{Fang1}.
\end{thebibliography}
\end{document}